\documentclass[aps,pre,twocolumn,superscriptaddress,noshowpacs]{revtex4}
\usepackage{amsmath}
\usepackage{amssymb}
\usepackage{tikz}
\usepackage{xcolor}
\usepackage{graphicx}
\usepackage{dcolumn}
\usepackage{bm}
\usepackage{epsfig}
\usepackage{psfrag}
\usepackage{latexsym}

\usepackage{dsfont,bbold,physics,xcolor}

\newcommand{\str}{\text{\,str\,}}
\newcommand{\re}{\text{\,Re\,}}
\newcommand{\im}{\text{\,Im\,}}
\newcommand{\sdet}{\text{sdet\,}}
\newcommand{\diag}{\text{diag\,}}
\newcommand{\1}{\mathds{1}}

\newcommand{\kb}{\mathbf{k}}
\newcommand{\refsec}[1]{\mbox{Sec.~\ref{#1}}}

\begin{document}

\title{Distribution of Off-Diagonal Cross Sections in Quantum Chaotic Scattering:\\
       Exact Results and Data Comparison}

     \author{Santosh Kumar} \email{skumar.physics@gmail.com}
     \affiliation{Department of Physics, Shiv Nadar University, Gautam
       Buddha Nagar, Uttar Pradesh 201314, India} \author{Barbara Dietz}
     \email{dietz@lzu.edu.cn} \affiliation{%
       School of Physical Science and Technology, and Key Laboratory
       for Magnetism and Magnetic Materials of MOE, Lanzhou
       University, Lanzhou, Gansu 730000, China } \author{Thomas Guhr}
     \email{thomas.guhr@uni-due.de} \affiliation{Fakult\"at f\"ur
       Physik, Universit\"at Duisburg-Essen, Lotharstra\ss{}e 1,
       D-47048 Duisburg, Germany} \author{Achim Richter}
     \email{richter@ikp.tu-darmstadt.de} \affiliation{Institut f\"{u}r
       Kernphysik, Technische Universit\"{a}t Darmstadt, D-64289
       Darmstadt, Germany}

\begin{abstract}
  The recently derived distributions for the scattering-matrix
  elements in quantum chaotic systems are not accessible in the
  majority of experiments, whereas the cross sections are. We
  analytically compute distributions for the off-diagonal cross
  sections in the Heidelberg approach, which is applicable to a wide
  range of quantum chaotic systems. We thus eventually fully solve a
  problem which already arose more than half a century ago in
  compound-nucleus scattering. We compare our results with data from
  microwave and compound-nucleus experiments, particularly addressing
  the transition from isolated resonances towards the Ericson regime
  of strongly overlapping ones.
\end{abstract}


\maketitle
\section{Introduction}
Scattering experiments are indispensable to understand the microscopic
world. Mainly developed in nuclear
physics~\cite{Ericson1963,Satchler1963,Brink1963,Porter1965,MW1969,Feshbach1993,Zelevinsky1996,GMW1998,PS2002,Weidenmueller2009},
scattering theory now finds various applications in condensed matter
physics~\cite{Lee1987,Jalabert1990,Beenakker1997,Alhassid2000}, 
in classical wave
systems~\cite{Weaver1989,Ellegaard1995,Gros2014}, in wireless
communication~\cite{CD2011} and other
fields~\cite{FKS2005,YAOA2010,MRW2010}.  An incoming wave in a
scattering channel $b$, say, is modified in the scattering zone,
\textit{e.g.}, by a nucleus as the target, and leaves it through a
scattering channel $a$. The elements $S_{ab}(E)$ of the associated
scattering matrix $S$ are complex numbers. They provide all
information on the changes in amplitude and phase, typically with
energy $E$. The $S$ matrix is unitary due to flux
conservation and its dimension coincides with the number $M$ of channels.
In a few cases both the modulus and the phase of the $S$-matrix elements can be measured directly,
\textit{e.g.}, in experiments with microwave cavities, microwave networks or reverberating
elastic objects~\cite{Stoeckmann1990,Graef1992,Hul2004,AEOGS2010}. In the majority of scattering experiments, particularly in quantum physics, the phase is not accessible. In mesoscopic quantum dots~\cite{Folk1996} the electron transport, that is, the conductance is measured instead, of which the fluctuations are well understood~\cite{Weidenmueller1990,Beenakker1997,Alhassid2000,Celardo2008}. In a scattering experiment involving quantum particles, {\it i.e.}, atoms~\cite{Main1992,Stania2005,Madronero2005,Frisch2014}, molecules~\cite{Reid1996,Mayle2013} or nuclei~\cite{Ericson1966}, only the incoming and outgoing particle current can be measured. Their ratio yields the cross sections. For $a\ne b$ they are given by 
\begin{equation}
 \sigma_{ab}(E) = |S_{ab}(E)|^2 = \left(\textrm{Re\,}S_{ab}(E)\right)^2 
                                      + \left(\textrm{Im\,}S_{ab}(E)\right)^2.
\label{crossab}
\end{equation}
This formula might have to be supplemented with multiplicative factors
of purely kinematic origin.

If the dynamics in the scattering zone is sufficiently complex or, in
a rather general sense, chaotic, scattering can usually be
thought of as a random process~\cite{Jung1997}.  There are in
principle two stochastic approaches to chaotic
scattering~\cite{Beenakker1997,GMW1998}. In the Mexico
approach~\cite{Mello1985,Mares2005}, the $S$ matrix as a whole is
viewed as a random matrix, whereas in the Heidelberg approach
randomness is assumed for the Hamiltonian $H$ describing the
internal dynamics in the interaction region. While the former has an
unrivaled conceptual elegance, the latter is better suited for
grasping important features of the internal dynamics since the
scattering process as such is fully modeled on the microscopic level.

We have three goals: First, we calculate within the Heidelberg approach the exact distribution of the
off-diagonal cross sections $\sigma_{ab}$ with $a\ne b$, corresponding to inelastic 
scattering or rearrangement collisions, thereby
providing the complete solution of a long-standing problem. It applies
from the regime of isolated resonances with average resonance
width $\Gamma$ smaller than the average resonance spacing $D$,
\textit{i.e.}, $\Gamma /D\ll 1$, all the way up to the Ericson
regime~\cite{Ericson1960} of strongly overlapping resonances, $\Gamma
/D\gg 1$. Second, we test our results by comparing with cross-section
data obtained in microwave and compound-nucleus experiments, focussing
on the transition to the Ericson regime.  Third, we provide a simple
and robust method to extract non-random contributions to the
cross-section distribution.

\section{Scattering matrix}
The Heidelberg approach~\cite{AWM1975,VWZ1985} is based on~\cite{MW1969}
\begin{align}
S_{ab}(E) &= \delta_{ab}-i 2\pi W_a^\dag G(E) W_b\, ,\\
G^{-1}(E) &= E\mathds{1}_N-H+i\pi\displaystyle\sum_{c=1}^M W_c W_c^\dag\, ,
\label{Sab}
\end{align}
where $G(E)$ is the matrix resolvent. The widths of the resonances generated by the poles of 
$G(E)$ in the complex energy plane exhibit non-trivial fluctuations~\cite{Fyodorov1997}.
They are controlled by the interplay between the Hamilton matrix $H$ describing the scattering zone
and the coupling vectors $W_c$ which account for the interaction between the channels
$c$ and the states of $H$.

Scattering can involve different time scales. In nuclear physics,
there are direct, non-random reactions on very short
time scales due to channel-channel coupling. On longer time scales a
compound nucleus is formed by the target and the incoming
particles. Its equilibration ensures a sufficient amount of
stochasticity, justifying the replacement of $H$ by a random
matrix. We assume absence of direct
coupling between the channels, implying that the coupling vectors
$W_c$ may be chosen orthogonal, $W_c^\dag W_d=\gamma_c
\delta_{cd}/\pi$~\cite{AWM1975,LW1991} where $\gamma_c$ is referred to
as partial width.  Depending on whether the system is time-reversal
invariant or noninvariant, $H$ either belongs to the Gaussian
Orthogonal Ensemble (GOE) or to the Gaussian Unitary Ensemble
(GUE)~\cite{Mehta2004,GMW1998} designated by the Dyson indices
$\beta=1$ and $\beta=2$, respectively.  The entries of the matrices
$H$ are Gaussian distributed, $\mathcal{P}(H)d[H] \sim
\exp\left(-\frac{\beta N}{4v^2}\text{tr}H^2\right)d[H]$ with variance
parameter $v^2$.  The flat measure $d[H]$ is the product of
differentials of all independent elements in the $N\times N$ matrix
$H$.  All physical quantities are measured on the local
scale of the mean level spacing. This implies universality,
\textit{i.e.}, a very large class of probability densities gives the
same result in the limit $N\to\infty$; see
Refs.~\cite{GMW1998,Mehta2004}.

\section{Cross-section distribution} Although the cross-section
distribution was of high interest already in the early days of
compound-nucleus and, more generally, of chaotic scattering, it
continued to resist an analytical solution~\cite{Ericson2016}. In a
seminal work using the supersymmetry method, Verbaarschot,
Weidenm\"uller and Zirnbauer~\cite{VWZ1985} derived the
exact two--point energy correlation function of the $S$-matrix
elements. Davis and Boos\'e calculated three-
and four-point correlation functions~\cite{DB1988,DB1989} and Fyodorov,
Savin and Sommers the distribution of the
diagonal $S$-matrix elements~\cite{FSS2005}. Rozhkov, Fyodorov, and
Weaver~\cite{RFW2003,RFW2004} computed a related quantity, namely the
statistics of transmitted power.  Putting forward a new variant of the
supersymmetry method, we recently calculated the distributions of the
real and the imaginary parts of the off--diagonal $S$
matrix~\cite{KN2013,NKSG2014}. In a related study, Fyodorov and Nock
obtained the distributions of off--diagonal elements of the Wigner $K$
matrix~\cite{FN2015}. Nevertheless, the cross-section distribution
remained out of reach, because the cross section (\ref{crossab})
depends on the real and imaginary parts of the $S$-matrix
element which are not independent. Thus, to compute 
it for $a\neq b$,
\begin{equation}
p(\sigma_{ab}) = \int_{-\infty}^{\infty}dx_1\int_{-\infty}^{\infty}dx_2 \,
         \delta(\sigma_{ab}-x_1^2-x_2^2)P(x_1,x_2) \ ,
\label{pcrs}
\end{equation}
the knowledge of the joint probability density function
\begin{equation}
P(x_1,x_2) = \int d[H] \mathcal{P}(H) \delta(x_1-\re S_{ab}) \delta(x_2-\im S_{ab})
\label{jpdfx}
\end{equation}
is inevitable. At first sight, one might expect that this task leads
to doubling the size of the supersymmetric non--linear sigma model as
compared to Refs.~\cite{FSS2005,KN2013,NKSG2014}, rendering further
evaluation forbiddingly complicated. However, we recently discovered
that a simple, yet far-reaching modification and generalization of our
supersymmetry technique in Refs.~\cite{KN2013,NKSG2014} yields $P(x_1,x_2)$
without enlarging this size.

\section{Joint probability density}
It turns out to be advantageous to employ the Fourier transform, \textit{i.e.},
the bivariate characteristic function
\begin{equation}
R(k_1,k_2) = \int d[H] \mathcal{P}(H) e^{-ik_1\re S_{ab}-ik_2\im S_{ab}}
\label{chark}
\end{equation}
in two dimensions, such that
\begin{equation}
P(x_1,x_2) = \frac{1}{4\pi^2}\int_{-\infty}^\infty dk_1\! \int_{-\infty}^\infty\! dk_2\, 
                  e^{ik_1 x_1+ik_2 x_2} R(k_1,k_2) \, .
\label{Px1x2}
\end{equation}
Anticpating the data analysis to follow we emphasize that the
characteristic function is obtained by sampling from the experimental
data as easily as the joint probability density itself.  With
Eq.~(\ref{Px1x2}) in Eq.~(\ref{pcrs}) and the complex variables
$\kb=k_1+ik_2$ and $\mathbf{x}=x_1+ix_2$, we find
\begin{equation} 
p(\sigma_{ab})=\frac{1}{4\pi^2}\int\! d^2 \mathbf{x} \int\! d^2\kb\,
            \delta(\sigma_{ab}-|\mathbf{x}|^2)e^{i \re(\kb^* \mathbf{x})}R(\kb) \ .
\label{psgmcomplex}
\end{equation}
The $\mathbf{x}$ integrals can be done with polar coordinates, 
\begin{equation}
p(\sigma_{ab}) = \frac{1}{4\pi} \int\!d^2\kb\, R(\kb) \,J_0\big(\sqrt{\sigma_{ab}}|\kb|\big) \ ,
\label{psgm1}
\end{equation}
expressing the cross-section distribution as a certain Bessel
transform of the characteristic function. The crucial step to make the
calculation of the latter feasible is to use
Eq.~(\ref{Sab}) in Eq.~(\ref{chark}) in the form
\begin{equation}
 R(\kb)= \int d[H] \mathcal{P}(H) \exp(-i\pi W^T A\, W) 
 \label{R1}
\end{equation}
with the $2N$ component vector $W^T=[W_a^T, W_b^T]$ for $a\neq b$, and
the $2N \times 2N$ Hermitian matrix
\begin{equation}
 A=\begin{bmatrix}
              0 & -i\kb^* G \\
              i\kb G^\dag & 0
   \end{bmatrix}
\label{adef}
\end{equation}
in terms of the resolvent in Eq.~(\ref{Sab}). In
Refs.~\cite{KN2013,NKSG2014} we proceeded similarly, but for marginal
distributions and thus univariate characteristic functions that depend
either on $k_1$ or on $k_2$. Absorbing them as complex variable $\kb$
into the definition of $A$ preserves its Hermiticity.
Hence, we may adjust all further steps in
Refs.~\cite{KN2013,NKSG2014} by moving the calculation into the
complex $\kb$ plane. We introduce bosonic integrals for a Fourier
transform of the characteristic function (\ref{R1}) in $W$ space to
invert the resolvent $G$ in $A$. A thereby occurring determinant is
written as a fermionic integral. This allows us to do the ensemble
average over the random matrices $H$ exactly.  We obtain a supermatrix
model that we bring onto the local scale by a saddlepoint
approximation for $N\to\infty$. This yields a supersymmetric nonlinear
sigma model extending the one in Refs.~\cite{KN2013,NKSG2014}. Details
are given in Sect.~I of the appendix~\refsec{suppl}.

For $\beta=2$ with unitarily invariant $H$ the final result for the
characteristic function is
\begin{align}
\label{RkUE}
\nonumber
R(\kb)=1-&\int_1^\infty d\lambda_1\int_{-1}^1 d\lambda_2 \frac{|\kb|^2}{4(\lambda_1-\lambda_2)^2}\,\mathcal{F}_\text{U}(\lambda_1,\lambda_2) \\
&~~\times \left(t_a^1 t_b^1+t_a^2 t_b^2\right)J_0\left(|\kb| \sqrt{t_a^1 t_b^1}\right) \ ,
\end{align}
with the channel factor
\begin{equation}
\label{Fu}
\mathcal{F}_\text{U}(\lambda_1,\lambda_2)
=\prod_{c=1}^M \frac{g_c^{+} + \lambda_2}{g_c^{+} + \lambda_1} \, ,
\end{equation}
where $t_c^j=\sqrt{|\lambda_j^2-1|}/(g_c^{+}+\lambda_j)$, and
$g_c{^\pm}=(v^2\pm\gamma_c^2)/(\gamma_c \sqrt{4v^2-E^2})$. The
parameter $g^{+}_c$ is related to the transmission coefficient or the
sticking probability $T_c=1-|S_{cc}|^2$ as $g^{+}_c=2/T_c-1$. The
remarkable fact that the characteristic function (\ref{RkUE}) depends
only on $|\kb|$ implies that the distribution of real and imaginary
parts of $S_{ab}$ are identical~\cite{KN2013,NKSG2014} for
$\beta=2$. For $\beta=1$ with orthogonally invariant $H$ we arrive at
\begin{align}
\label{RkOE}
\nonumber
&R(\kb)=1+\frac{1}{8\pi}\int_{-1}^1 d\lambda_0 \int_1^\infty d\lambda_1\int_1^\infty d\lambda_2 \int_0^{2\pi}d\psi  \\
&\times \mathcal{J}(\lambda_0,\lambda_1,\lambda_2)\mathcal{F}_\text{O}\left(\lambda_0,\lambda_1,\lambda_2\right)\left(\kappa_1+\kappa_2+\kappa_3+\kappa_4\right) \, .
\end{align}
The Jacobian in the above expression is given by
\begin{equation}
\mathcal{J}=\frac{(1-\lambda_0^2)|\lambda_1-\lambda_2|}
{2(\lambda_1^2-1)^{1/2}(\lambda_2^2-1)^{1/2}(\lambda_1-\lambda_0)^2(\lambda_2-\lambda_0)^2}\, ,
\end{equation}
and the channel factor reads
\begin{equation}
\label{Fo}
\mathcal{F}_\text{O}(\lambda_0,\lambda_1,\lambda_2)
=\prod_{c=1}^M \frac{g_c^{+} + \lambda_0}
{(g_c^{+} + \lambda_1)^{1/2}(g_c^{+} + \lambda_2)^{1/2}}\, .
\end{equation}
The $\kappa$'s in Eq.~\eqref{RkOE} depend on $g^{\pm}_c$ and the
complex $\kb$ in a nontrivial way; see~\refsec{suppl}. 

\section{Comparison with microwave data}
\begin{figure}[ht!]
\centering
\includegraphics[width=0.8\linewidth]{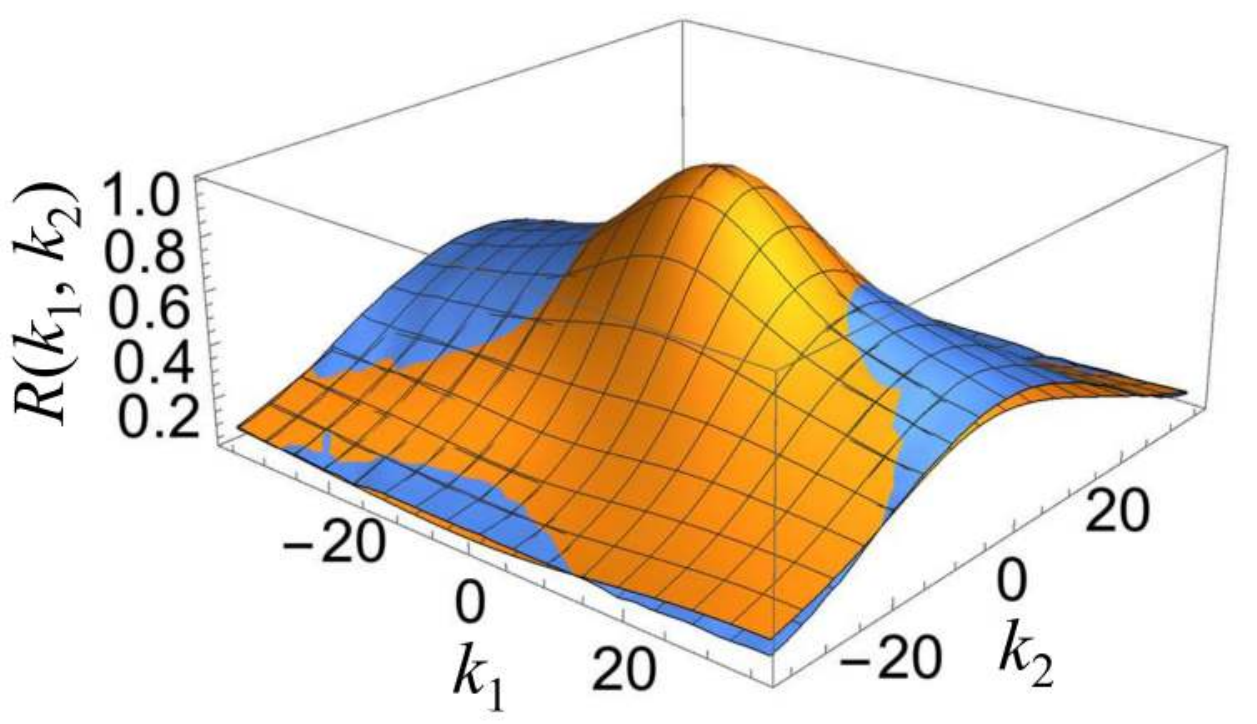}
\caption{Bivariate characteristic function $R(k_1,k_2)$ in the
  frequency range 10-11~GHz. Analytical result (blue) and
  microwave data (orange).}
\label{Rk1k2_10-11GHz}
\end{figure}

\begin{figure}
\centering
\includegraphics[width=0.8\linewidth]{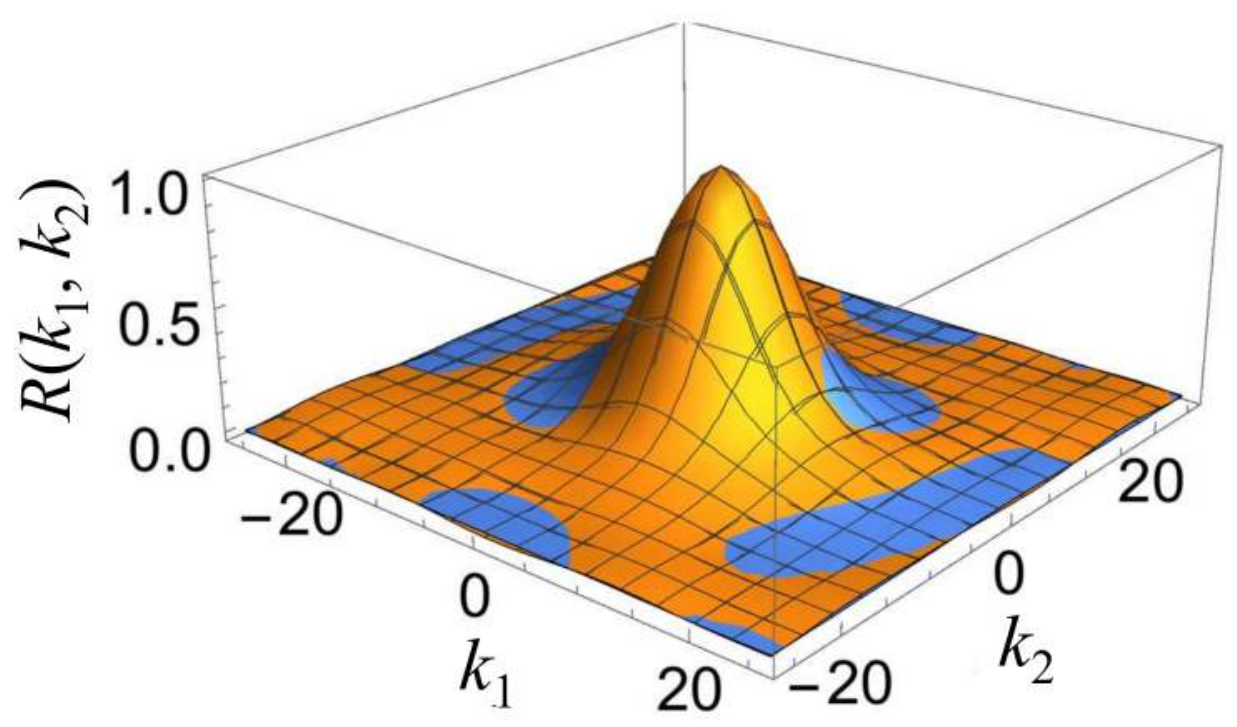}
\caption{As Fig.~\ref{Rk1k2_10-11GHz}, but in the frequency range
  24-25~GHz.}
\label{Rk1k2_24-25GHz}
\end{figure}

\begin{figure}[ht!]
\includegraphics[width=0.9\linewidth]{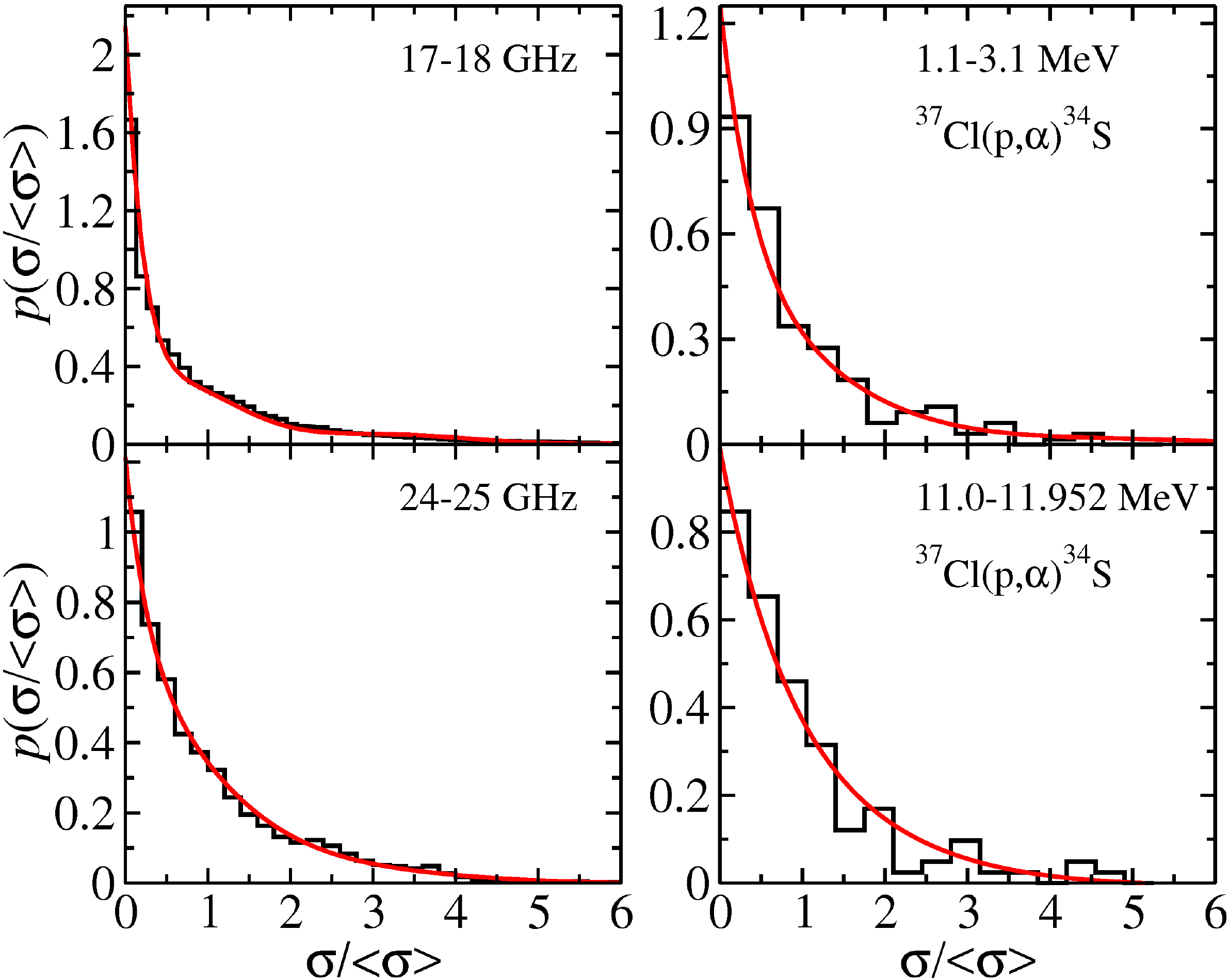}
    \caption{Distribution of normalized cross sections. Experimental
      data as histograms from microwave (left) and nuclear experiments
      (right), respectively.  Analytical results as solid red lines.}
\label{Sigma_Distr}
\end{figure}

The mathematical equivalence of spectra of two--dimensional
quantum billiards and flat microwave resonators is used to
experimentally explore a variety of quantum chaotic phenomena in
closed~\cite{Stoeckmann1990,Sridhar1991,Graef1992,Dietz2015} and open
systems~\cite{KMMS2005,KSW2005,H2005,Hul2005,L2008,Dietz2009,Dietz2010a}.
Here, we use the data measured for a microwave billiard in the shape
of a classically chaotic tilted--stadium billiard; see 
Refs.~\cite{Dietz2009,Dietz2010a,Dietz2010b} for
experimental details.  The $S$-matrix elements $S_{ab}$ were
measured in steps of 100 kHz in a range from 1 to 25~GHz.  Their
fluctuation properties were evaluated in frequency windows of 1~GHz to
guarantee a negligible secular variation of the coupling vectors
$W_c$. In Ref.~\cite{KN2013} we analyzed the marginal distributions of
real and imaginary parts of $S_{ab}$ and the corresponding univariate
characteristic functions separately.  We now compare our new
analytical results for the joint probability density $P(x_1,x_2)$, for
the bivariate characteristic function $R(k_1,k_2)$ and for the cross
section distribution $p(\sigma_{ab})$ with these data.  
Figure~\ref{Rk1k2_10-11GHz} shows the bivariate characteristic
function in the frequency range 10-11~GHz. Plotted are
the analytical and experimental results together. The same comparison is shown in
Fig.~\ref{Rk1k2_24-25GHz} for the frequency range 24-25~GHz; see also~\refsec{suppl}.
The agreement is very good in both cases. For the lower frequencies,
the peak is broad and heavy--tailed, corresponding to a non-Gaussian
joint probability density. For the higher ones, the peak is narrow and
Gaussian--like, yielding the joint probability density with a
nearly Gaussian shape for the frequency range 24-25~GHz, 
displayed in Fig.~\ref{Px1x2_24-25GHz}.  To explain these
results, we point out that the system undergoes with increasing frequency 
a transition from isolated resonances to largely
overlapping ones, \textit{i.e.}, to the onset region of the
Ericson regime~\cite{Dietz2010b}.  For
the frequency ranges 10-11~GHz and 24-25~GHz, we have
$\Gamma/D=0.23$ and $\Gamma/d=1.21$, respectively.  In the Ericson
regime, scattering matrices and cross sections are random functions and
the peaks in the spectra cannot be associated with particular
resonances, implying that the distribution of the $S$-matrix
elements is Gaussian~\cite{Brink1963,AWM1975}. According to
Eq.~(\ref{pcrs}), the distribution of normalized cross sections is then exponential
with $p(0)=1$. To test this, we also compare in Fig.~\ref{Sigma_Distr}
our results for the distribution of cross sections normalized to their
mean with the data. As seen, our exact results compare well to all
regimes including the transition region.  The nearly exponential form
with $p(0)>1$ in the frequency range 24-25~GHz clearly indicates that
we are in the onset of the Ericson regime.
\begin{figure}[ht!]
\includegraphics[width=0.8\linewidth]{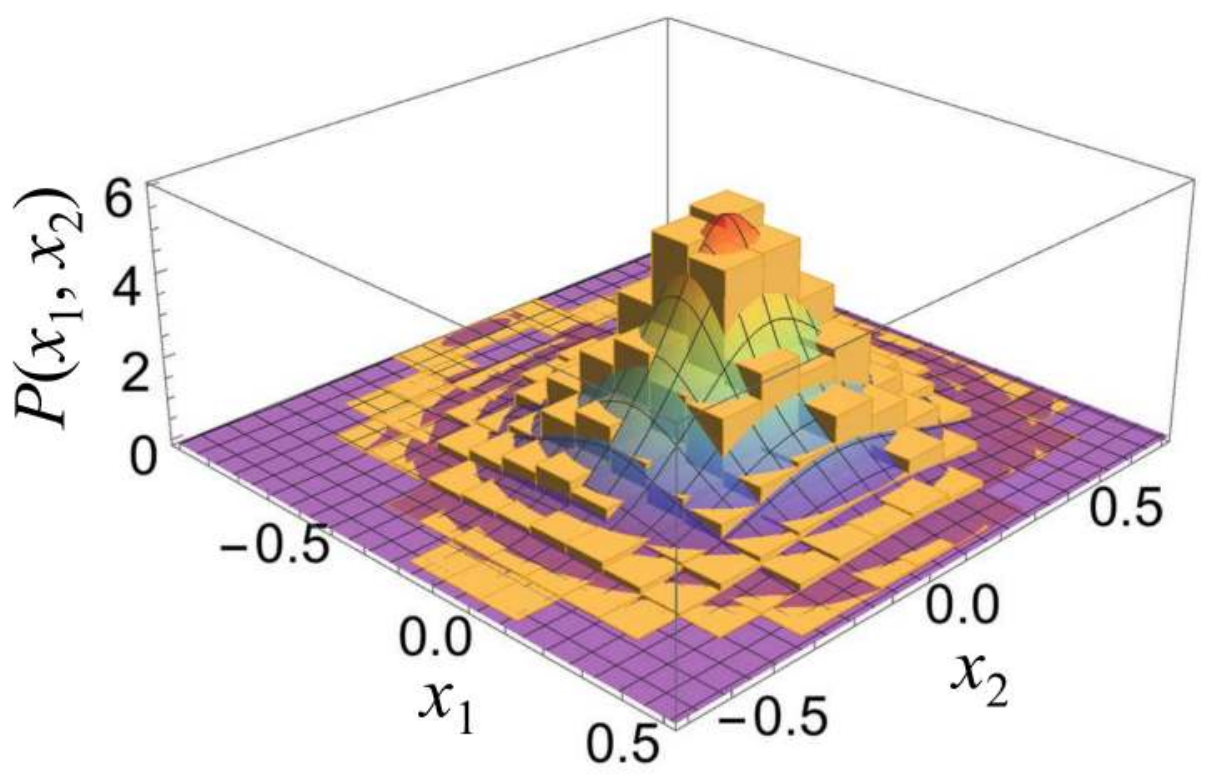}
\caption{Joint probability density $P(x_1,x_2)$, analytical (surface)
and microwave data (histogram) in the frequency range 24-25~GHz.}
\label{Px1x2_24-25GHz}
\end{figure}

\section{Comparison with compound-nucleus data}
\begin{figure}
\includegraphics[width=1.0\linewidth]{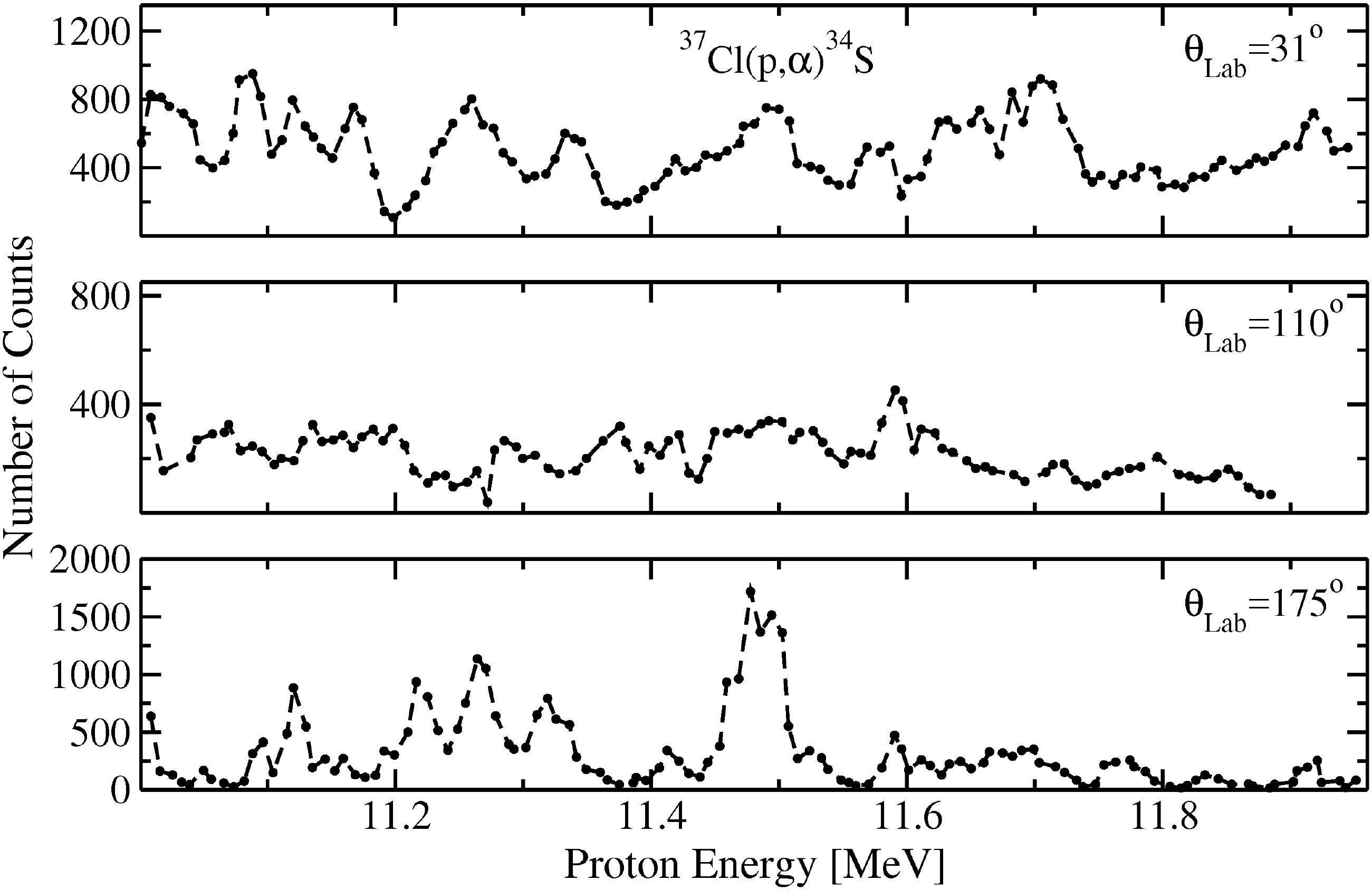}
\caption{Excitation functions for the reaction
  $^{37}\rm{Cl(p,}\alpha)^{34}\rm{S}$ in the Ericson regime for
  scattering angles $31^\circ,\, 110^\circ$ and $175^\circ$ from top
  to bottom. Digitized from Ref.~\cite{WBMR1966}.}
\label{kerndaten1}
\end{figure}

We also use data from historical measurements of the compound-nuclear
reaction
$^{37}\rm{Cl(p,}\alpha)^{34}\rm{S}$~\cite{vonBrentano1964,WBMR1966,Richter1966}. In
Ref.~\cite{WBMR1966}, excitation functions were measured in steps of
8~keV in the proton-energy range 11-11.952~MeV for 12 scattering
angles between $31^\circ$ to $175^\circ$.  Importantly, these data are
fully in the Ericson regime with $\Gamma/D \approx 27-36$.  In
Fig.~\ref{kerndaten1} we show a selection of three such excitation
functions for $31^\circ,\, 110^\circ$ and $175^\circ$. At smaller
angles, one observes a background, \textit{i.e.}, a nonzero minimum
value of the excitation function. It is due to direct reactions, in
which, \textit{e.g.}, an incoming particle kicks out an $\alpha$
particle without formation of a compound nucleus. As such processes
are stronger in forward than in backward direction, the background
disappears at larger angles. In addition, they are barely affected by
the chaotic dynamics in the scattering zone and thus cannot be
random. Hence, their energy dependence is marginal and we may safely
substract the background to obtain the fluctuating compound-nuclear 
contribution. In Fig.~\ref{Sigma_Distr} we compare the
distribution of normalized cross-sections, \textit{i.e.},
$p(\sigma/\langle\sigma\rangle)$ obtained from the $175^\circ$
measurement with the analytical prediction. For this, we use $M=5$
(effective) open channels and all transmission coefficients $T_c=0.99$
in accordance with Ref.~\cite{WBMR1966}, leading to an exponential. We
find a very good match.

\begin{figure}
\includegraphics[width=0.8\linewidth]{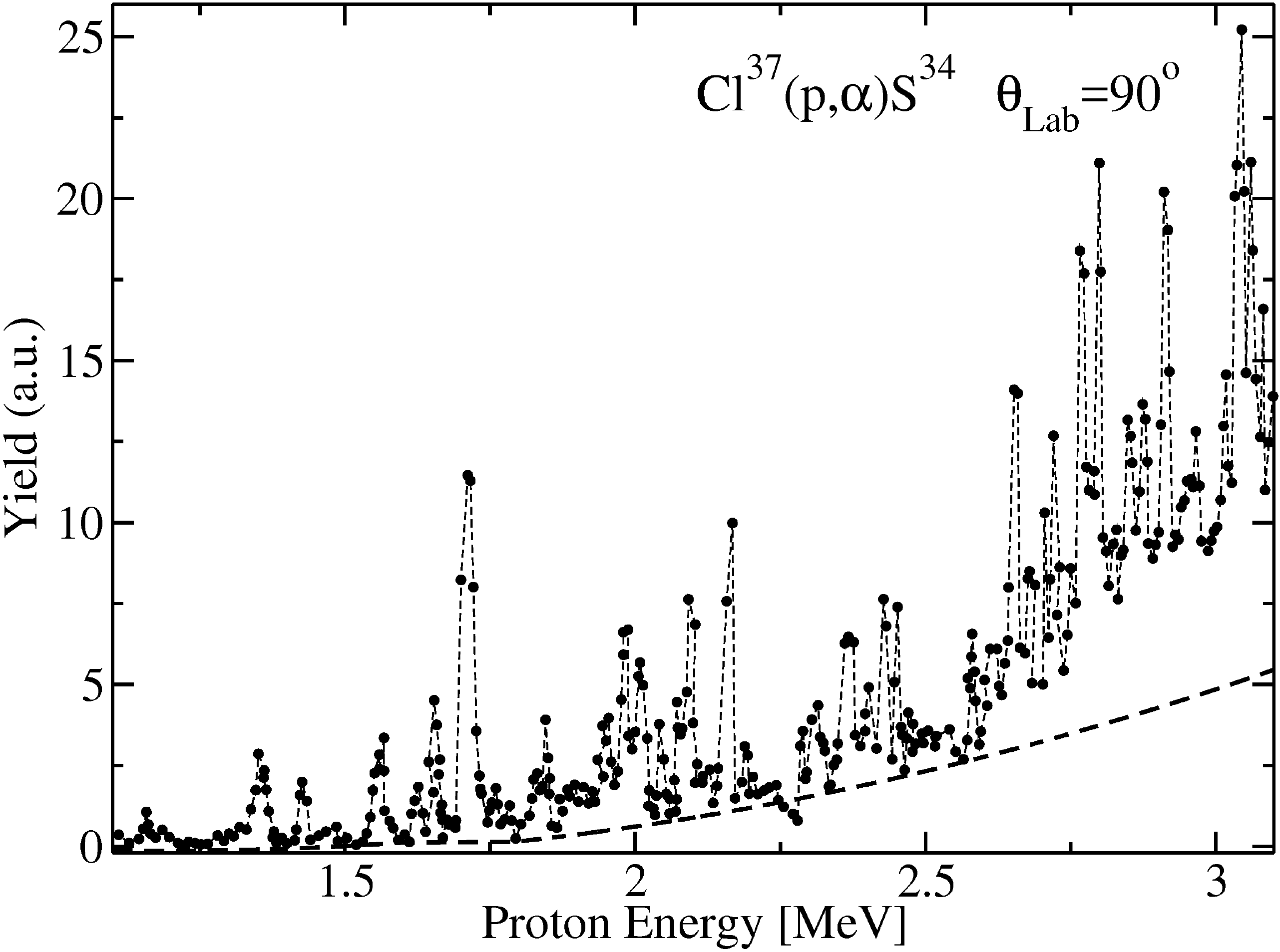}
\caption{Excitation functions for the reaction
  $^{37}\rm{Cl(p,}\alpha)^{34}\rm{S}$ in below the Ericson regime for
  scattering angle $90^\circ$. Digitized from
  Ref.~\cite{Clarke1959}.}
\label{kerndaten2}
\end{figure}

To complete our studies we, furthermore, apply our analytical results
to nuclear data in the region of weakly overlapping resonances. In
Ref.~\cite{Clarke1959}, the reaction
$^{37}\rm{Cl(p,}\alpha)^{34}\rm{S}$ was measured in the proton-energy
range 1.1-3.1~MeV at a scattering angle of $90^\circ$. These data,
shown in Fig.~\ref{kerndaten2}, exhibit an unusually sharp increase at
an energy of approximately $2.6$~MeV which is due to experimental
imperfections. We thus restrict the data analysis to the energy range
1.1-2.6~MeV.  The background stemming from the direct reactions is
a smoothly increasing function of energy, hence
subtracting it is more involved than in the previously considered
case; see Fig.~\ref{kerndaten1}. This reflects a general problem in
analyzing compound-nuclear data.  Unfortunately, we cannot exploit
recent progress that has been made employing the $K$
matrix~\cite{KTW2015,KC2016}, since it relies on the knowledge of the
$S$-matrix elements.  Instead, we put forward a seemingly new
empirical method which is based on the observation that the
peak exhibited by the cross-section distribution of compound-nuclear reactions 
at $\sigma=0$ is shifted to a nonzero value by direct contributions. Thus, we fit the excitation function
below 2.6~MeV with a second-order polynomial, which we then subtract from the data.
This leads to the experimental cross-section distribution displayed in
Fig.~\ref{Sigma_Distr} which is peaked at zero. Our analytical result
is very well capable of describing this clearly non-exponential
distribution for $M=10$ effective channels and $T_c=0.7$.

\section{Conclusions}
We solved a long-standing problem by exactly calculating the
distribution of the off-diagonal cross sections within the Heidelberg
approach. This facilitates, for the first time, an analysis of
distributions for the large number of systems, in which only the cross
sections can be measured. We performed a detailed comparison with
microwave and nuclear data, focusing on the transition from the regime
of isolated resonances towards the Ericson regime. Our analytical
results describe the data very well in all regimes. We are not aware
of any comparable study for distributions and characteristic
functions. In the course of our data comparison, we came up with a
seemingly new and robust method to substract the direct part in
cross-section data which only relies on experimental information.

\begin{acknowledgments}
  We are grateful to A.~Nock for his help in the initial stages of this
  project, and to P. von Neumann-Cosel who helped us to find 
  Ref.~\cite{Clarke1959}. We acknowledge fruitful discussions with 
  T.~Kawano and H.A.~Weidenm\"uller. SK also
acknowledges the support by the grant EMR/2016/000823 provided by SERB,
DST, Government of India. This work was supported by the Deutsche
  Forschungsgemeinschaft (DFG) within the Collaborative Research
  Centers 634 and 1245.
\end{acknowledgments}

\begin{widetext}
\section{Appendix}\label{suppl}
\appendix
\section{Derivation of the characteristic functions for the cases $\beta=1,2$\label{SectI}}
Although the derivation to follow can be, with the necessary
adjustments, inferred from Refs.~\cite{KN2013, NKSG2014}, we document
here the crucial steps for the readers less experienced with
supersymmetry calculations.  For the case of unitarily invariant $H$,
i.e. $\beta=2$, in order to bring the matrix $A$ into a diagonal form, we consider
the following transformations:
\begin{equation}\label{beta2transf}
 z \to \Xi_{z} z, \quad  z^\dag \to z^\dag, \quad \zeta \to \Xi_{\zeta} \zeta, \quad  \zeta^\dag \to \zeta^\dag
\end{equation}
with
\begin{align}
 \Xi_{z} = \begin{bmatrix}
               0 & -i\kb^*  \\
               -i\kb  & 0 
              \end{bmatrix} \otimes \1_N \, ,
~~
 \Xi_{\zeta} = \begin{bmatrix}
                0 & i\kb^* \\
                -i\kb & 0 
              \end{bmatrix} \otimes 1_N \, .
\end{align}
The characteristic function, therefore, can be written as
\begin{equation}
 R(\kb) = (-1)^N\! \int\! d[H]\, \mathcal{P}(H)\! \int\! d[\Psi] e^{\frac{i}{2}( \mathbf{U} ^\dag\Psi+ \Psi^\dag \mathbf{W})+\frac{i}{4\pi} \Psi^\dag \mathcal{A}^{-1} \Psi } \, ,
\end{equation}
where 
\begin{equation}
\mathcal{A}^{-1}=
\diag[-(G^{-1})^\dag,G^{-1},-(G^{-1})^\dag,-G^{-1}]
\end{equation}
is a block diagonal $4N \times 4N$ matrix which is independent of $\kb$. Also 
\begin{equation}
\mathbf{U}^T = [i \kb^* W_b^T, i \kb W_a^T, 0, 0],\ \mathbf{W}^T =
(1/2)[W_a^T, W_b^T, 0, 0]
\end{equation}
are $4N$-dimensional vectors composed of
the coupling vectors. Apart from the expressions of $\mathbf{U}$ and
$\mathbf{W}$, the characteristic function for the joint probability
distribution $\mathcal{P}(H)$ has the same form as that for the joint probablity distribution of the real and imaginary parts of $S$, compare with Eq. (11) 
of Ref.~\cite{KN2013} and Eq.~(21) of Ref.~\cite{NKSG2014}. In particular, the definition of
$\mathcal{A}^{-1}$ is exactly the same. This allows us to proceed 
as in Refs.~\cite{KN2013, NKSG2014}, leading to an exact
representation of the characteristic function for arbitrary $N$ in
terms of a matrix integral in superspace:
\begin{align}
\label{exact-beta2}
\nonumber
 &R(\kb) = \int\! d[\tau] e^{-\frac{4\pi^2 N}{v^2} \str \tau^2 -\frac{i}{4} \mathbf{U}^\dag (\mathbf{L}^{1/2}\mathcal{T} \mathbf{L}^{1/2})^{-1} \mathbf{W}} \sdet\!^{-1}\,\mathcal{T} \ , \\
 \nonumber
 &\mathcal{T} = \tau_E \otimes \1_N +\frac{i}{4} L \otimes \sum_{c=1}^M W_c W_c^T, \quad \tau_E=\tau-\frac{E}{4\pi} \1_4 \ , \\
 &L=\diag(+1,-1,+1,-1), \quad \mathbf{L}=L \otimes \1_N \ .
\end{align}
Here ``str'' and ``sdet'' denote the supertrace and superdeterminant,
respectively. The supermatrix integral can be performed using
saddle-point analysis in the $N\rightarrow \infty$ limit, thereby
mapping it to a nonlinear sigma model. By employing the standard
parametrization as in~\cite{KN2013,NKSG2014}, we obtain the final
expression Eq.~(13) for the characteristic function in the $\beta=2$ case.

We now focus on the orthogonally invariant $H$, {\it i.e.}, $\beta=1$ and
instead of the transformation given in Eq.~(18) we
implement the following one:
\begin{equation}
 z \to \Xi_{z} z,~~  z^\dag \to z^\dag \Xi_{z^\dag}, ~~ \zeta \to \Xi_{\zeta} \zeta,~~   \zeta^\dag \to \zeta^\dag \Xi_{\zeta^\dag}
\label{Def_zeta}
\end{equation}
with
\begin{align*}
 &\Xi_{z} = \begin{bmatrix}
               0 & \sqrt{-i\kb^*}  \\
               \sqrt{-i\kb}  & 0 
              \end{bmatrix} \otimes \1_N  \ ,
\\
 & \Xi_{z^\dag} = \begin{bmatrix}
                0 & \sqrt{2i\kb^*} \\
                \sqrt{-2i\kb} & 0 
              \end{bmatrix} \otimes \1_N \ ,
\\
 &\Xi_{\zeta} = \begin{bmatrix}
                     \sqrt{-i\kb} & 0 \\
                     0 & \sqrt{-i\kb^*}
                    \end{bmatrix} \otimes \1_N \ ,
\\
 & \Xi_{\zeta^\dag} = \begin{bmatrix}
                           \sqrt{-2i\kb} & 0 \\
                           0 & \sqrt{2i\kb^*}
                         \end{bmatrix} \otimes \1_N \ .
\end{align*}
We furthermore consider the supervector $\Psi^T=[x_a^T, y_a^T,
x_b^T,y_b^T, \zeta_a^T, \zeta_a^\dag,\zeta_b^T, \zeta_b^\dag]$, where
we decomposed $z=x+iy$ into its real and imaginary part. This yields for the
characteristic function
\begin{equation}
  R(\kb) = (-1)^N \int\! d[H]\, \mathcal{P}(H)\! \int\! d[\Psi] e^{i \Psi^\dag \mathbf{V}} e^{\frac{i}{4\pi} \Psi^\dag  \mathcal{A}^{-1} \Psi } \ ,
\end{equation}
where 
\begin{equation}
\mathcal{A}^{-1}=
\diag[-(G^{-1})^\dag,G^{-1},-(G^{-1})^\dag,-G^{-1}] \otimes {\1_2}
\end{equation}
is a block diagonal $8N \times 8N$ matrix which is independent of $\kb$,
and 
\begin{equation}
\mathbf{V}^T= (1/2)[\sqrt{-i\kb} (W_a+W_b)^T, -\sqrt{-i\kb}
(W_a-W_b)^T, \sqrt{-i\kb^*} (W_a+W_b)^T, i\sqrt{-i\kb^*} (W_a-W_b)^T,
0, 0, 0, 0] 
\end{equation}
is a $\kb$-dependent $8N$-dimensional vector composed of
the coupling vectors.  It should be noted that in order to render
$\mathcal{A}^{-1}$ block-diagonal and independent of $\kb$, already
transforming $z$ and $\zeta$ omitting the square-roots in the
transformation-matrices $\Xi_z(\kb), \Xi_{\zeta}(\kb)$ in Eq.~(\ref{Def_zeta}) would have
been sufficient. However, transforming also $z^\dag$, $\zeta^\dag$
along with $z$, $\zeta$ has the advantage that the ensuing expression
for $\mathbf{V}$ becomes simpler and numerically stable.  Apart from
the expression of $\mathbf{V}$, the characteristic function for the
joint probability distribution has, as in the case $\beta=2$, exactly
the same form as that for the joint probability distribution of the 
real- or imaginary part of $S$, compare
with Eq.~(12) of Ref.~\cite{KN2013} and Eq.~(70) of Ref.~\cite{NKSG2014}. In
particular, the definition of $\mathcal{A}^{-1}$ is exactly the
same. This allows us to perform the same computations as in the
previous works, leading to an exact representation of the
characteristic function for arbitrary $N$ in terms of a matrix
integral in superspace,
\begin{align}
\label{exact-beta1}
\nonumber
 &R(\kb) = \int\! d[\tau] e^{-\frac{4\pi^2 N}{v^2} \str \tau^2 -\frac{i}{4} \mathbf{V}^T (\mathbf{L}^{1/2}\mathcal{T} \mathbf{L}^{1/2})^{-1} \mathbf{V}} \sdet\!^{-1/2}\,\mathcal{T} \ , \\
\nonumber
 &\mathcal{T} = \tau_E \otimes \1_N +\frac{i}{4} L \otimes \sum_{c=1}^M W_c W_c^T, \quad \tau_E=\tau-\frac{E}{4\pi} \1_8, \\
 &L=\diag(+1,-1,+1,-1) \otimes \1_2, \quad \mathbf{L}=L \otimes \1_N \ .
\end{align}
Here $\tau$ is an $8 \times 8$ dimensional supermatrix of appropriate
symmetry. Again, considering the large $N$ limit and using the standard
parametrization for the supermatrix and performing the steps outlined
in \cite{NKSG2014}, we obtain the final result Eq.~(15). 
In the top figures of Fig.~\ref{Rk1k2_17-18GHz}  we compare our new analytical results for the bivariate characteristic function $R(k_1,k_2)$ with those for microwave data in the frequency ranges 10-11~GHz (left), 17-18~GHz (middle) and 24-25~GHz (right). The lower figures show the differencse between the analytical and the experimental results.
In Fig.\ref{Px1x2_24-25GHz} we compare the corresponding analytical and experimental results for the joint probability density $P(x_1,x_2)$ in the frequency ranges 17-18~GHz (left) and 24-25~GHz (right). 
\begin{figure}[ht!]
\centering
\includegraphics[width=1.0\linewidth]{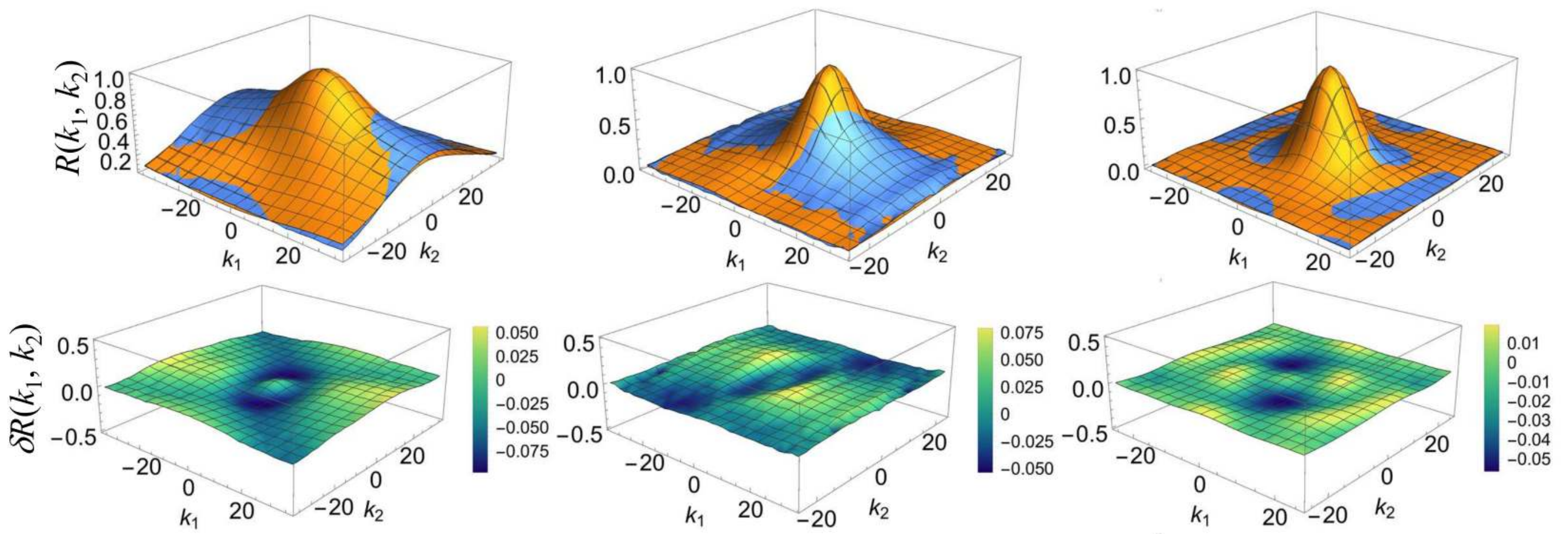}
\caption{Bivariate characteristic function $R(k_1,k_2)$ obtained from the microwave data~\cite{Dietz2009,Dietz2010a,Dietz2010b} in the frequency ranges 10-11~GHz (left), 17-18 GHz (middle) and 24-25 GHz (right). Top: Analytical result (blue) and microwave data (orange). Bottom: Difference between the two.}
\label{Rk1k2_17-18GHz}
\end{figure}
\begin{figure}[ht!]
\centering
\includegraphics[width=0.8\linewidth]{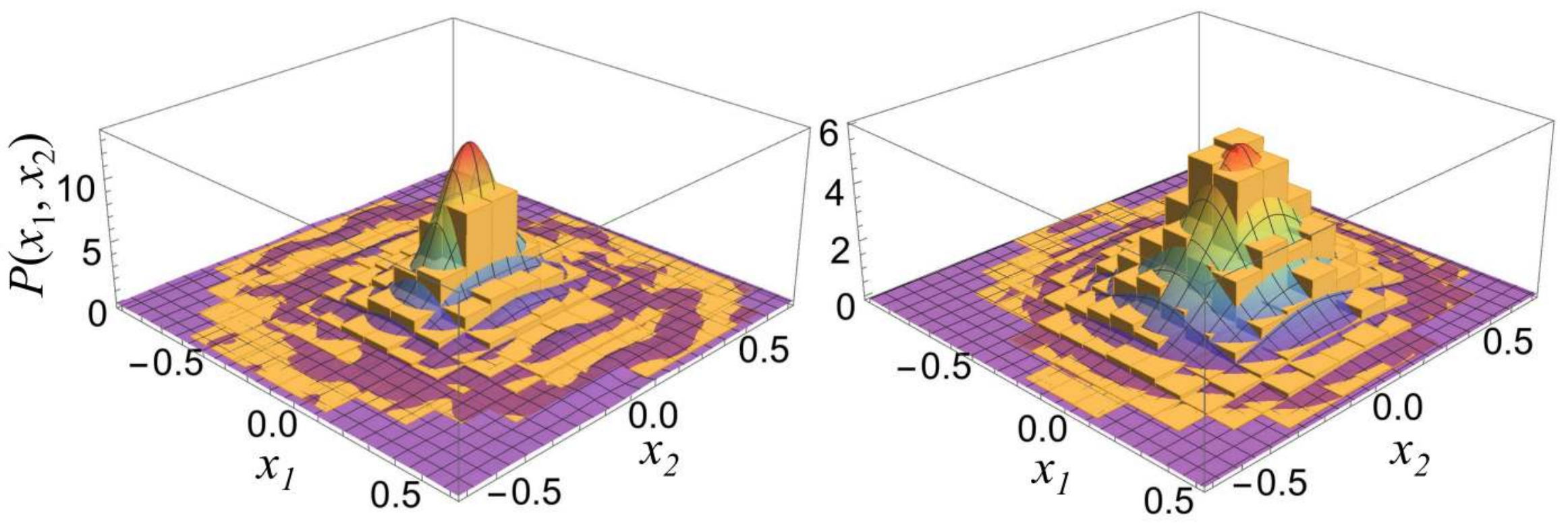}
\caption{Joint probability density $P(x_1,x_2)$, analytical (surface) and microwave data (histogram) in the frequency ranges 17-18 GHz (left) and 24-25 GHz (right).}
\label{Px1x2_24-25GHz}
\end{figure}

\section{Definition of the $\kappa$'s\label{SectII}}
To define the $\kappa$'s appearing in Eq.~(22) we need the following:
\begin{equation}
p_c^j=\frac{|\mathbf{k}|}{8}\frac{\sqrt{|\lambda_j^2-1|}}{(g_c^{+} +\lambda_j)},~~~~ j=0,1,2,
\end{equation}
\begin{equation}
p_c^{\pm}=p_c^1\pm p_c^2,
\end{equation}
\begin{equation}
\label{qcp}
q_c^{+}=\frac{\mathbf{k}}{8i}\left(\frac{E}{\sqrt{4v^2-E^2}}+i g_c^{-}\right)\left(\frac{1}{g_c^{+}+\lambda_1}+\frac{1}{g_c^{+}+\lambda_2}
-\frac{2}{g_c^{+}+\lambda_0}\right),
\end{equation}
\begin{equation}
\label{qcm}
q_c^{-}=\frac{\mathbf{k}}{8i}\left(\frac{E}{\sqrt{4v^2-E^2}}+i g_c^{-}\right)\left(\frac{1}{g_c^{+}+\lambda_1}-\frac{1}{g_c^{+}+\lambda_2}\right),
\end{equation}
\begin{equation}
\label{rcp}
r_c^{+}=\frac{i\mathbf{k}^*}{8}\left(\frac{E}{\sqrt{4v^2-E^2}}-i g_c^{-}\right)\left(\frac{1}{g_c^{+}+\lambda_1}+\frac{1}{g_c^{+}+\lambda_2}
-\frac{2}{g_c^{+}+\lambda_0}\right),
\end{equation}
\begin{equation}
\label{rcm}
r_c^{-}=\frac{i\mathbf{k}^*}{8}\left(\frac{E}{\sqrt{4v^2-E^2}}-i g_c^{-}\right)\left(\frac{1}{g_c^{+}+\lambda_1}-\frac{1}{g_c^{+}+\lambda_2}\right).
\end{equation}
It is to be noted that $r_c^{\pm}=(q_c^{\pm})^*$. We also define the quantities $l=X/Y$, $m=Y/X$, $\omega=2\sqrt{X Y}$, where
\begin{equation}
X=2p_a^{+}+q_a^{-}e^{-i2\psi}+r_a^{-}e^{i2\psi},~~~~
Y=2p_b^{+}+q_b^{-}e^{i2\psi}+r_b^{-}e^{-i2\psi}.
\end{equation}
It can be verified that $\omega^2$ is real for all the values of parameters involved and assumes the values from 0 to 1. The $\kappa$'s are given as
\begin{eqnarray}
\nonumber
 \kappa_1=\kappa_{11} J_1(\omega),
~~~ 
\kappa_2=\kappa_{21} J_0( \omega)+\kappa_{22} J_2( \omega),\hspace{1.8cm}\\
 \kappa_3=\kappa_{31} J_1(\omega)+\kappa_{32} J_3(\omega),
~~~
 \kappa_4=\kappa_{41} J_0(\omega)+\kappa_{42} J_2(\omega)+\kappa_{43} J_4(\omega).
\end{eqnarray}
~\\
The coefficients with the Bessel functions above are as follows
\begin{equation}
\kappa_{11}=-(9/8)\{p_a^{+}m^{1/2} \}_{+},
\end{equation}
\begin{eqnarray}
\kappa_{21}=-(1/4)(128 p_a^0p_b^0+14p_a^{+}p_b^{+}+32p_a^{-}p_b^{-})
-\{3e^{i2\psi}(p_a^{-}q_b^{+}+p_b^{-}r_a^{+})\}_{+}
-\{e^{-4i\psi}q_a^{-}r_b^{-}\}_{+},
\end{eqnarray}
\begin{equation}
\kappa_{22}=-(1/4)\{(p_a^{+}p_a^{+}-4q_a^{-}r_a^{-})m\}_{+}, 
\end{equation}
\begin{eqnarray}
\nonumber
&&\kappa_{31}=\big\{-2\big[(p_a^{+}p_a^{+}+q_a^{-}r_a^{-})m^{1/2}+2(8p_a^0p_b^0+p_a^{+}p_b^{+}+p_a^{-}p_b^{-})l^{1/2}\big](e^{i2\psi}q_b^{-}+e^{-i2\psi}r_b^{-})\\
\nonumber
&&+2\big[(p_a^{+}p_b^{-}+4p_a^{-}p_b^{+})m^{1/2}+p_b^{+}p_b^{-}l^{1/2}\big](e^{-i2\psi}q_a^{+}+e^{i2\psi}r_a^{+})+
\big[16p_a^0(2p_a^0 p_b^{+}-3p_b^0 p_a^{+})\\
\nonumber
&&
+6p_a^{+}(q_a^{+} q_b^{+}+r_a^{+} r_b^{+})+2p_b^{+}(4q_a^{+} r_a^{+}-q_a^{-} r_a^{-})-4p_a^{-}(p_a^{+}p_b^{-}-2p_a^{-}p_b^{+})-3p_a^{+}(q_a^{-}q_b^{-}+r_a^{-}r_b^{-}+p_a^{+}p_b^{+})
\\
\nonumber
&&-(e^{-i4\psi}/2)q_a^{-}(4p_a^{+}r_b^{-}+3p_b^{+}q_a^{-}+2e^{-i2\psi}q_a^{-}r_b^{-}-8e^{i2\psi}r_a^{+}r_b^{+})\\
\nonumber
&&-(e^{i4\psi}/2)r_a^{-}(4p_a^{+}q_b^{-}+3p_b^{+}r_a^{-}+2e^{i2\psi}q_b^{-}r_a^{-}-8e^{-i2\psi}q_a^{+}q_b^{+})\big]m^{1/2}\big\}_{+},
\end{eqnarray}
\begin{eqnarray}
 \kappa_{32}=\big\{p_a^{+}\big[(p_a^{+}p_a^{+}+2q_a^{-}r_a^{-})
 +(3/2)(e^{-i4\psi} q_a^{-}q_a^{-}+e^{i4\psi}r_a^{-}r_a^{-})
 +(2p_a^{+}p_a^{+}+q_a^{-}r_a^{-})(e^{-i2\psi} q_a^{-}+e^{i2\psi}r_a^{-})\big]m^{3/2}\big\}_{+},
\end{eqnarray}
\begin{eqnarray}
\nonumber
 \kappa_{41}\!\!\!&=&\!\!\!32 \big[2 p_a^0 p_a^0(p_b^{-}+e^{i2\psi}q_b^{+})(p_b^{-}+e^{-i2\psi}r_b^{+})
   +2 p_b^0 p_b^0(p_a^{-}+e^{-i2\psi}q_a^{+})(p_a^{-}+e^{i2\psi}r_a^{+})\\
     \nonumber
   &+&\!\!\! p_a^0 p_b^0\big((p_a^{+}+e^{-i2\psi}q_a^{-})(p_b^{+}+e^{-i2\psi}r_b^{-})+(p_a^{+}+e^{i2\psi}r_a^{-})(p_b^{+}+e^{i2\psi}q_b^{-})\big)\big]\\
   \nonumber
   &+&\!\!\! 256 p_a^0 p_a^0p_b^0 p_b^0
   +(p_a^{+}+e^{-i2\psi}q_a^{-})^2(p_b^{+}+e^{-i2\psi}r_b^{-})^2
   +(p_a^{+}+e^{i2\psi}r_a^{-})^2(p_b^{+}+e^{i2\psi}q_b^{-})^2\\
    \nonumber
   &+&\!\!\! 4\big[(p_a^{+}+e^{-i2\psi}q_a^{-})(p_b^{+}+e^{i2\psi}q_b^{-})
-2(p_a^{-}+e^{-i2\psi}q_a^{+})(p_b^{-}+e^{i2\psi}q_b^{+})\big]\\
   &&\times\big[(p_a^{+}+e^{i2\psi}r_a^{-})(p_b^{+}+e^{-i2\psi}r_b^{-})
-2(p_a^{-}+e^{i2\psi}r_a^{+})(p_b^{-}+e^{-i2\psi}r_b^{+})\big],
\end{eqnarray}
\begin{eqnarray}
\nonumber
 \kappa_{42}=-32p_a^0p_b^0\big[(p_a^{+}+e^{-i2\psi}q_a^{-})(p_a^{+}+e^{i2\psi}r_a^{-})m+(p_b^{+}+e^{i2\psi}q_b^{-})(p_b^{+}+e^{-i2\psi}r_b^{-}) l ]\hspace{2cm}\\
 \nonumber
+2\big[(p_a^{+}+e^{-i2\psi}q_a^{-})(p_b^{+}+e^{i2\psi}q_b^{-})
-2(p_a^{-}+e^{-i2\psi}q_a^{+})(p_b^{-}+e^{i2\psi}q_b^{+})\big]
\big[(p_a^{+}+e^{i2\psi}r_a^{-})^2 m
 +(p_b^{+}+e^{-i2\psi}r_b^{-})^2 l\big]\\
 + 2\big[(p_a^{+}+e^{i2\psi}r_a^{-})(p_b^{+}+e^{-i2\psi}r_b^{-})
-2(p_a^{-}+e^{i2\psi}r_a^{+})(p_b^{-}+e^{-i2\psi}r_b^{+})\big]
\big[(p_a^{+}+e^{-i2\psi}q_a^{-})^2 m
 +(p_b^{+}+e^{i2\psi}q_b^{-})^2 l \big],
\end{eqnarray}
\begin{eqnarray}
 \kappa_{43}=\big\{(p_a^{+}+e^{-i2\psi}q_a^{-})^2(p_a^{+}+e^{i2\psi}r_a^{-})^2 m^2\big\}_{+}.
\end{eqnarray}
In the above equations,  an expression $\mathcal{E}$ involving $a,b,l,m,\psi$ enclosed in the bracket $\{\,\}_{\pm}$ represents $\{\mathcal{E}(a,b,l,m,\psi)\}_{\pm}
:=\mathcal{E}(a,b,l,m,\psi)\pm\mathcal{E}(b,a,m,l,-\psi)$.
\end{widetext}
\end{document}